\documentclass[11pt]{amsart}
\usepackage{amssymb}
\newtheorem{Lemma}{Lemma} [section]
\newtheorem{Proposition}{Proposition} [section]

\newtheorem{Corollary}{Corollary} [section]

\newtheorem{Example}{Example} [section]
\newtheorem{Remark}{Remark} [section]
\newtheorem{Conjecture}{Conjecture}[section]
\def\proof{\par{\it Proof}. \ignorespaces}
\def\endproof{{\ \vbox{\hrule\hbox{%
   \vrule height1.3ex\hskip0.8ex\vrule}\hrule }}\par}
\newenvironment{Proof}{\proof}{\endproof}

\begin{document}

\title{ Topology of the real part of hyperelliptic
Jacobian associated with the periodic Toda lattice}

\author{ Yuji Kodama$^{\dagger}$ \\ }

\thanks{${\dagger}$)  Department of Mathematics, Ohio State University,
Columbus, OH 43210\endgraf
{\it E-mail address\/}: kodama@@math.ohio-state.edu}

\begin{abstract}
This paper concerns the topology of the isospectral {\it real} manifold of the
${\mathfrak sl}(N)$ periodic Toda lattice consisting of $2^{N-1}$
different systems. The solutions of those systems contain blow-ups, and
the set of those singular points defines a devisor of the manifold.
Then adding the divisor, the manifold is compactified as the real part
of the $(N-1)$-dimensional Jacobi variety associated with a hyperelliptic
Riemann surface of genus $g=N-1$. We also study the real structure of
the divisor, and then provide conjectures on the topology of the affine part
of the real Jacobian and on the gluing rule over the divisor to compactify the
manifold based upon the sign-representation of the Weyl group of ${\mathfrak sl}(N)$.

\end{abstract}

\maketitle

\markboth{YUJI KODAMA}
  {TOPOLOGY OF PERIODIC TODA}

\section{Preliminary}
\renewcommand{\theequation}{1.\arabic{equation}}\setcounter{equation}{0}
\renewcommand{\thefigure}{1.\arabic{figure}}\setcounter{figure}{0}

Let us start with an infinite Toda lattice which is defined as a hamiltonian
system with the hamiltonian,
\begin{equation}
\nonumber
H=\displaystyle{{1\over 2}\sum_{k\in {\Bbb Z}}
P_k^2 +\sum_{k\in {\Bbb Z}} \epsilon_k\exp\left(Q_k-Q_{k+1}\right)},
\end{equation}
where $\epsilon_k\in \{\pm 1\}$.
Because of the signs $\epsilon_k$ in the potential term, the
solution blows up if at least one sign is negative \cite{kodama:98}.
With Flaschka's coordinates, $
a_k:=\epsilon_k\exp\left(Q_k-Q_{k+1}\right)$ and $
b_k:= -P_k$, Hamilton's equation with the hamiltonian $H$ is given by
\begin{equation}
\label{toda}
\left\{
\begin{array}{ll}
\displaystyle{{da_k\over dt}}&=a_k(b_{k+1}-b_k), \\
\displaystyle{{db_k\over dt}}&=a_k-a_{k-1}.
\end{array}
\right.
\end{equation}

\begin{Remark}
The sign $\epsilon_k$ is invariant under the Toda flow, and the affine part of
the isospectral manifold of the Toda flow consists of $2^{N-1}$ disconnected
pieces labeled by the signs of $a_k$'s (note $\prod_{k=1}^N a_k$ is invariant).
\end{Remark}

The infinite Toda lattice in $\{(a_k,b_k): k\in {\Bbb Z}\}$ can be put in
a matrix equation, the Lax formulation,
\begin{equation}
{dL \over dt}=[L,A],
\end{equation}
where $L$ is an $\infty \times\infty$ matrix given by
\begin{eqnarray}
\label{L}
L = \left(
\begin{matrix}
\ddots & \ddots & \ddots & \ddots   &       &       &   &   &  &\\
 \ddots    & a_{-1} & b_0   & 1   & 0     &       &   &   &  &\\
    & 0      & a_0   & b_1 & 1     &0   &   &   &  &\\
{}     &        &\ddots &\ddots &\ddots &\ddots & \ddots  &   &  &\\
       &        &       &\ddots &\ddots &\ddots &\ddots &\ddots & &\\
 &      &        &       &0      & a_{k-2} & b_{k-1} & 1 &0 & \\
  &     &        &       &       &0        & a_{k-1} & b_k & 1 & \ddots  \\
   &    &        &       &       &         & \ddots  &\ddots&\ddots &\ddots
\end{matrix}
\right)
\end{eqnarray}
and $A$ is given by the strictly lower triangular part of $L$, i.e.
$A=(L)_{-}$.
The Lax equation is just the compatibility
condition of the following linear equations
under isospectral condition,
\begin{equation}
\left\{
\begin{array}{ll}
L\phi &=\lambda \phi , \\
\displaystyle{{d\phi \over dt}}&=A\phi .
\end{array}
\right.
\end{equation}

An ${\mathfrak sl}(N)$ periodic Toda lattice is then given by the conditions,
$ Q_{k+N}=Q_k$ and $P_{k+N}=P_k$ which imply
\begin{equation}
\label{ab}
a_{k+N}=a_k, \quad b_{k+N}=b_k.
\end{equation}
Then the eigenvector $\phi=(\cdots, \phi_0, \phi_1, \phi_2, \cdots)^T$
satisfies
\begin{equation}
\label{phiperiod}
\phi_{k+N}=z\phi_k,
\end{equation}
where $z$ is the spectral parameter which is the eigenvalue of the monodromy
of this periodic problem. The Lax matrix for the ${\mathfrak sl}(N)$ periodic
Toda lattice is now given by
\begin{eqnarray}
\label{LP}
L_P = \left(
\begin{matrix}
b_1 & 1 & 0 & \cdots & a_Nz^{-1} \\
a_1 & b_2 & 1 & \cdots & 0 \\
\vdots & \ddots & \ddots & \ddots & \vdots \\
0 & \cdots & a_{N-2} & b_{N-1} & 1 \\
z & \cdots & \cdots & a_{N-1} & b_N \\
\end{matrix}
\right) .
\end{eqnarray}
The characteristic equation for $L_P$ defines the algebraic curve,
\begin{equation}
\label{characteristicLP}
{\rm det}(L_P-\lambda I)=-\displaystyle{\left(
z+{\prod_{i=1}^N a_i \over z}-P(\lambda)\right)}=0
\end{equation}
which is a
1-dimensional affine variety on $(\lambda,z) \in {\Bbb C}^2$. Here $P(\lambda)$
is an $N$-th polynomial of $\lambda$ given by
\begin{equation}
\nonumber
P(\lambda):=\Delta_{1,N}(\lambda)-a_N\Delta_{2,N-1}(\lambda),
\end{equation}
where $\Delta_{m,n}(\lambda)$ for $n\ge m$ is defined by
\begin{eqnarray}
\label{Deltamn}
\Delta_{m,n}(\lambda) := \left|
\begin{matrix}
b_m-\lambda & 1 & 0 & \cdots & 0 \\
a_m & b_{m+1}-\lambda & 1 & \cdots & 0 \\
\vdots & \ddots & \ddots & \ddots & \vdots \\
0 & \cdots & a_{n-2} & b_{n-1}-\lambda & 1 \\
0 & \cdots & 0 & a_{n-1} & b_{n}-\lambda \\
\end{matrix}
\right| .
\end{eqnarray}
We assume that the discriminant of (\ref{characteristicLP}),
$F(\lambda):=P(\lambda)^2-4\prod_{j=1}^Na_j$, has all distinct roots, i.e.
\begin{equation}
\label{D}
F(\lambda)=\displaystyle{\prod_{k=1}^{2N_R}(\lambda-\lambda_k)
\prod_{l=1}^{N_I}\Big((\lambda-\nu_l)(\lambda-{\bar\nu}_l)\Big)}.
\end{equation}
with all different roots $\lambda_j\in {\Bbb R},\nu_l\in{\Bbb C}$, and
$N_R+N_I=N$.
This implies that
the algebraic curve having this condition is a smooth affine variety.
We say that a matrix $L_P$ is in the class (P) if
the corresponding periodic matrix $L$ satisfies this condition.
The phase space for the periodic lattice is given by
\begin{equation}
\nonumber
Z_{\Bbb R}^{P}:=\Big\{(a_1,\cdots,a_N,b_1,\cdots,b_N)\in{\Bbb R}^{2N}:
L_P ~{\rm ~is~in~the~class~(P)} \Big\}~.
\end{equation}

Since the ${\mathfrak sl}(N)$ Toda lattice is a hamiltonian system of degree
$N$, its integrability can be shown by finding $N$ independent and
involutive integrals. Those integrals $I_k$ are given by
the coefficients of the characteristic equation (\ref{characteristicLP})
where $P(\lambda)$ is defined as
\begin{equation}
\label{P}
P(\lambda)=(-1)^N\left(\lambda^N+\sum_{k=1}^N(-1)^kI_k(L_P)\lambda^{N-k}
\right),
\end{equation}
and there is an additional integral obtained from the residue with respect to the
spectral parameter $z$, $\prod_{i=1}^N a_i$, i.e.
\begin{equation}
\nonumber
\left\{
\begin{array}{ll}
I_1(L_P)&=~\displaystyle{\sum_{i=1}^N b_i},\\
I_2(L_P)&=~\displaystyle{\sum_{i>j} b_ib_j-\sum_{i=1}^Na_i}, \\
 & \cdots \\
 I_N(L_P) &=~\displaystyle{\Delta_{1,N}(0)-a_N\Delta_{2,
N-1}(0)=\prod_{i=1}^Nb_i+\cdots,} \\
 I_{N+1}(L_P)&=~\displaystyle{\prod_{i=1}^Na_i}.
\end{array}
\right.
\end{equation}
Then the isospectral set is defined by
\begin{equation}
\nonumber
Z_{\Bbb R}^P(\gamma):=\Big\{(a_1,\cdots,a_N,b_1,\cdots,b_N)\in Z_{\Bbb R}^P:
I_k(L_P)=\gamma_k\in{\Bbb R},~k=1,\cdots,N+1\Big\}
\end{equation}
which is the affine part of the compactified manifold ${\hat Z}_{\Bbb
R}^P(\gamma)$ of ${\rm dim}Z^P_{\Bbb R}(\gamma)=N-1$,
and with a divisor ${\Theta}$ associated to the blow-ups of
$a_k, k=1,\cdots,N$, we have \cite{adler:91}, 
\begin{equation}
\label{ZR}
Z_{\Bbb R}^P(\gamma)={\hat Z}_{\Bbb R}^P(\gamma)\setminus { \Theta},
\quad {\rm with} \quad {\Theta}=\bigcup_{k=1}^N\{a_k^{-1}=0\}.
\end{equation}
It turns out that the compact manifold ${\hat Z}_{\Bbb R}^P(\gamma)$
can be identified as the real part of the Jacobian
${\Bbb C}^g/\Gamma$ with the lattice $\Gamma$ defined by the period matrix
$\Omega$ associated with the Riemann surface of genus $g$ (see the next
section for
the details). The Riemann surface
is determined by the spectral curve, and the genus is just $g=N-1$.
We will discuss some details of this statement in the following sections.
The main purpose in this paper is to find the topological structure
of the divisor $\Theta$ in the real part of the Jacobian and to determine
the gluing rule on
the divisor to
compactify the affine variety $Z_{\Bbb R}^P(\gamma)$.
To study this, we start with a brief review of algebraic curve and
the solution of periodic Toda lattice in the next section.

\section{Algebraic curves and periodic Toda lattices}
\renewcommand{\theequation}{2.\arabic{equation}}\setcounter{equation}{0}
\renewcommand{\thefigure}{2.\arabic{figure}}\setcounter{figure}{0}
\subsection{Riemann surface and Jacobian}
The characteristic equation ${\rm det}(L_P-\lambda I)=0$ determines
a real plane curve ${\mathcal C}$ (1-dimensional affine curve on ${\Bbb C}^2$);
\begin{equation}
\label{y}
y^2=P(\lambda)^2-4=\displaystyle{\prod_{k=1}^{2N}(\lambda-\lambda_k)},
\end{equation}
where we have made a coordinate change from $(\lambda,z)$ to
$(\lambda, y)$ with $y=2z+P(\lambda)$. Also the complex conjugate pairs of the roots are denoted as
$(\lambda_{N_{R}+2k-1}=\nu_k, \lambda_{N_{R}+2k}={\bar \nu}_k)$
for $k=1,\cdots,N_I$ as in (\ref{D}).
Then a compact Riemann surface associated with the plane curve ${\mathcal C}$
is obtained by adding points at infinity by introducing a second chart
${\mathcal C}'$;
\begin{equation}
\nonumber
 (y')^2=\displaystyle{
\prod_{k=1}^{2N}(1-\lambda_k\lambda')}
\end{equation}
which is glued to the chart $\mathcal C$ with the isomorphism,
$\lambda'=1/\lambda, y'=y/ \lambda^{N}$.
This Riemann surface ${\mathcal R}={\mathcal C}\cup{\mathcal C}'$ is called
a hyperelliptic Riemann surface of genus $g=N-1$,
and it is topologically
a compact surface with $g$ handles. On the $\mathcal R$ one can
define a morphism $\pi:{\mathcal R}\to {\Bbb C}P^1$ by $(\lambda,y)\mapsto
\lambda$ on the chart $\mathcal C$ and similarly on the chart ${\mathcal C}'$.
Note that the morphism $\pi$ is a two-fold covering map on ${\Bbb C}P^1$
except at the branch points $\lambda_k$'s.
The points at $\infty$ of $\mathcal R$, which are denoted as
$p_{\infty}^{\pm}$,
are given by the points $(\lambda'=0,y'=\pm 1)$, i.e.
\begin{equation}
{\mathcal R}={\mathcal R}_0\cup \{p_{\infty}^+,p_{\infty}^-\}.
\end{equation}
where the ${\mathcal R}_0$ is the affine part of $\mathcal R$.
On the ${\mathcal R}$, we also define
a symplectic basis of $H_1({\mathcal R},{\Bbb Z})$ denoted as
$\{(\alpha_j,\beta_j):
j=1,\cdots,g\}$, in which the paths $\alpha_j$'s are disjoint from
each other as are the paths $\beta_j$'s, and $\alpha_i,\beta_j$ intersect
only if $i=j$ and then in one point.

Now let us define the {\it real} part of the Riemann surface,
denoted as ${\mathcal R}_{\Bbb R}$.  It is defined as the set of
fixed points of the antiholomorphic involution obtained by conjugation,
$(\lambda,y)\mapsto ({\bar \lambda},{\bar y})$,
\begin{equation}
\nonumber
{\mathcal R}_{\Bbb R}:=\Big\{(\lambda,y)\in {\mathcal R}: \lambda={\bar
\lambda},~
y={\bar y}\Big\}.
\end{equation}
Then it is easy to see:
\begin{Proposition}
\label{rankrealR}
Let $\mathcal R$ be a compact (smooth) Riemann surface of genus $g$.
 Then the real part ${\mathcal R}_{\Bbb R}$ is isomorphic to $N_R$
 disjoint union of circles, ${\Bbb R}P^1\cong S^1$,
\begin{equation}
\nonumber
{\mathcal R}_{\Bbb R}\cong \overbrace{S^1\sqcup \cdots \sqcup S^1}^{N_R}
\quad ({\rm
disjoint~union}),
\end{equation}
where $2N_R$ is the number of real zeros in (\ref{y}), i.e. the
real ramification points of $\mathcal R$.
\end{Proposition}

On the Riemann surface ${\mathcal R}$ of genus $g$, one can also have $g$
holomorphic differentials (the Riemann-Roch theorem), which we denote
$\omega_1,\cdots,\omega_g$. We normalize those differentials with
\begin{equation}
\label{normalization}
\oint_{\alpha_i}\omega_j=\delta_{i,j}, \quad {\rm for} ~ i,j=1,\cdots,g.
\end{equation}
Then the Jacobian ${\rm Jac}({\mathcal R})$
associated with the ${\mathcal R}$ is defined as a complex torus
with a lattice $\Gamma$ of rank $2g$;
\begin{equation}
\nonumber
{\rm Jac}({\mathcal R})={\Bbb C}^g/\Gamma, \quad
\Gamma:=\{\Omega m : m=(m_1,\cdots,m_{2g})^t\in {\Bbb Z}^{2g}\},
\end{equation}
where $\Omega$ is a $g\times 2g$ matrix, the period matrix, defined by
\begin{equation}
\label{periodicmatrix}
\displaystyle{\Omega=\left(
I~B \right), \quad {\rm with}\quad
B=\left(\oint_{\beta_i}\omega_j\right)_{1\le i,j\le g},}
\end{equation}
where $I$ is the $g\times g$ identity matrix.
It turns out that our purpose of studying the topological structure
of the isospectral manifold of periodic Toda lattice is just to study the
topological structure of the real part of the Jacobian and its
divisors corresponding to the blow-ups of the Toda flow.

\subsection{Periodic Toda lattice}
The Toda flow can be described as a linear flow on the Jacobian
${\rm Jac}(\mathcal R)$ of the Riemann surface $\mathcal R$. Let us first
illustrate the solution of a simplest case with
$N=2$ ($g=1$), and the general case will be discussed in the next
section.

The Lax matrix $L_P$ in this case is given by
\begin{equation}
\nonumber
L_P=\left(
\begin{matrix}
b_1 & 1+a_2z^{-1} \\
z+a_1 & b_2
\end{matrix}
\right)
\end{equation}
whose spectral curve defines an elliptic curve,
\begin{equation}
\nonumber
{\rm det}(L_P-\lambda I)=-\displaystyle{\left(z+{a_1a_2\over
z}-P(\lambda)\right)}=0.
\end{equation}
The polynomial $P(\lambda)$ is given by
\begin{equation}
\label{g1I2}
P(\lambda)=\lambda^2+I_2, \quad {\rm with} \quad I_2=-b_1^2-a_1-
\displaystyle{{1\over a_1}},
\end{equation}
where we have assumed $I_1=b_1+b_2=0$ and $I_3=a_1a_2=1$.
Then the affine part of the isospectral manifold is given by
\begin{equation}
\nonumber
Z_{\Bbb R}^P(\gamma)=\Big\{(a_1,b_1)\in {\Bbb R}^2:
b_1^2+a_1+{1\over a_1}=\gamma_2\in{\Bbb R}\Big\}
\cong \left\{
\begin{matrix}
 S^1\sqcup {\Bbb R}\sqcup {\Bbb R} \quad & {\rm if} ~~ \gamma_2>2 , \\
 {\Bbb R}\sqcup {\Bbb R} \quad & {\rm if} ~~ \gamma_2<2 .
\end{matrix}
\right.
\end{equation}
Thus we have
\begin{equation}
\label{g1Jac}
Z_{\Bbb R}^P(\gamma)\cong
{\mathcal R}_{\Bbb R}\setminus \{p_{\infty}^+, p_{\infty}^-\}=
{\mathcal R}_{\Bbb R}\cap {\mathcal R}_{0}.
\end{equation}
Two disconnected pieces of $\Bbb R$ are compactified to make a circle $S^1$,
that is, the corresponding solution blows up once in each point $p_{\infty}^+$
or $p_{\infty}^-$ and total twice in one cycle.
Thus
the compactified manifold ${\hat Z}_{\Bbb R}^P(\gamma)$ is just ${\mathcal
R}_{\Bbb R}$ (see below for more precise statement).

The Toda flow is determined by the equations,
\begin{equation}
\nonumber
\displaystyle{{da_1\over dt}=-2a_1b_1, \quad {db_1\over dt}=a_1-{1\over a_1}}.
\end{equation}
whose solution can be obtained by elliminating $a_1$ from (\ref{g1I2}),
\begin{equation}
\nonumber
\displaystyle{{db_1\over dt}=\pm \sqrt{(b_1^2+I_2)^2-4},}.
\end{equation}
This can be also written in the form,
\begin{equation}
\nonumber
\displaystyle{{d\lambda \over y(\lambda)}\Big|_{\lambda=b_1}=\pm dt},
\end{equation}
Here $y(\lambda)=\sqrt{(\lambda^2+I_2)^2-4}$, and the differential
$\omega_1={d\lambda\over y}$ is the holomorphic differential on $\mathcal R$.
Then the solution can be written in the {\it elliptic} integral,
\begin{equation}
\label{g1sol}
\displaystyle{\int_{b_0}^{b_1}\omega_1=\pm t + \delta_1}
\end{equation}
where $b_0$ is a fixed point, and $\delta_1$ is an arbitrary constant.
Since the integral (\ref{g1sol}) depends on the path from $b_0$
to $b_1$, it should be defined
on the Jacobian ${\rm Jac}({\mathcal R})$,
a complex torus ${\Bbb C}/\Gamma$ with the lattice $\Gamma$ generated by the
period matrix $\Omega=(1~B_{11})$ with $B_{11}=\oint_{\beta_1}\omega_1$.
Furthermore the elliptic integral (\ref{g1sol}) should be on the real
part of the Jacobian.
Namely we have the following maps with an Abel-Jacobi map $\upsilon$ to describe
the Toda flow;
\begin{equation}
\nonumber
\begin{matrix}
Z_{\Bbb R}^P(\gamma) &\longrightarrow & {\mathcal R}_{\Bbb R}
 & \overset{\upsilon}{\longrightarrow}
&{\rm Jac}({\mathcal R})_{\Bbb R}\\
   b_1 &\longmapsto & \mu_1 &\longmapsto &
\Big[\int_{\mu_*}^{\mu_1}\omega_1\Big]
 \end{matrix}
 \end{equation}
 where $\mu_*$ is a fixed point on ${\mathcal R}_{\Bbb R}$, the real
 part of the Riemann surface.
This implies that the Toda flow is {\it linear} on the Jacobian,
and the solution $b_1(t)$ is given by solving the
Jacobi inversion problem (just the Jacobi elliptic function
in this case), which is defined as the inverse map of $\upsilon$;
\begin{equation}
\nonumber
\psi:{\rm Jac}({\mathcal R}) \longrightarrow {\mathcal R}.
\end{equation}
Then the solution of the Jacobi inversion problem is obtained by the
zeros of Riemann theta function (see the next section for an explicit
form of the solution of Toda flow for genus $g>1$).
Now the question is to determine the structure of the real part of
${\rm Jac}({\mathcal R})\setminus \Theta$ which is isomorphic to
$Z_{\Bbb R}^P(\gamma)$.
In this example, the $\Theta$ is given by
\begin{equation}
\nonumber
\Theta=\{\upsilon(p_{\infty}^+),\upsilon(p_{\infty}^-)\},
\end{equation}
and the compactified manifold is given by ${\hat Z}_{\Bbb R}^P(\gamma)
\cong {\mathcal R}_{\Bbb R}\cong{\rm Jac}({\mathcal R})_{\Bbb R}$, i.e.
\begin{equation}
{\rm Jac}({\mathcal R})_{\Bbb R}\cong
\left\{
\begin{matrix}
S^1\times {\Bbb Z}/2{\Bbb Z}, \quad &{\rm for}~~\gamma_2>2, \\
S^1 , \quad &{\rm for}~~ \gamma_2<2 .\\
\end{matrix}
\right.
\end{equation}
where each $S^1$ is labeled by the signs $(\epsilon_1,\epsilon_2)$, i.e.
$(++)$ or $(--)$ for $\gamma_2>2$, and $(--)$ for $\gamma_2<2$.
Thus one of $S^1$ for $\gamma_2>2$ is obtained by gluing two pieces of
$\Bbb R$
on the $\Theta$. Finding the gluing structure of the affine pieces
of the real part of the Jacobian, ${\rm Jac}({\mathcal R})_{\Bbb R}
\setminus \Theta\cong Z_{\Bbb R}^P(\gamma)$ is one of the main purpose of
the present study.
\vskip 0.5cm

For the case of arbitrary genus, it has been known for quite a while that
the number of connected components
in the real part of the Jacobian,
denoted as $n({\rm Jac}_{\Bbb R}):=
{\rm rank}H_0({\rm Jac}({\mathcal R})_{\Bbb R},{\Bbb Z})$,
is completely determined by
the number of connected ones in the real part of Riemann suface,
$N_R={\rm rank}
H_0({\mathcal R}_{\Bbb R},{\Bbb Z})$, that is, we have:
\begin{Proposition}
\label{numberofconn}
(\cite{weichold:83}, see also \cite{audin:96, gross:81})
\label{realJac}
Let ${\mathcal R}$ be a genus $g$ compact Riemann surface with non-empty
real component, $N_R\ne 0$.
Then each connected component of ${\rm Jac}({\mathcal R})_{\Bbb R}$ is
isomorphic to
$g$-dimansional real torus, ${\Bbb T}^g={\Bbb R}^g/{\Bbb Z}^g$, and
\begin{equation}
\nonumber
{\rm Jac}({\mathcal R})_{\Bbb R}\cong {\Bbb T}^g\times \left({\Bbb Z}/2{\Bbb
Z}\right)^m,
\end{equation}
where $m=N_R-1\le g$, and thus $n({\rm Jac}_{\Bbb R})=2^m$.
\end{Proposition}
We will give the proof based on the solution structure of the periodic Toda
lattice in the next section.
Then we will study the topological
structure of
the affine parts of ${\rm Jac}({\mathcal R})_{\Bbb R}$ and the divisor
$\Theta$,
 and find the gluing rule among the affine parts labeled by
 the signs $(\epsilon_1,\cdots,\epsilon_N)$ for the compactification.
 It was shown in \cite{adler:91} that the divisor $\Theta$ consists of
 $g+1$ irreducible components $\Theta_k$ which corresponds to the blow-up
 of $a_k$, (see (\ref{ZR})).
 We then expect that each torus ${\Bbb T}^g$ in Proposition \ref{realJac} is
divided into
 $g+1$ affine varieties through those divisors $\Theta_k$.
 In order to study the case with arbitrary genus, we will give a brief review of
 the solutions of periodic Toda lattice in the next section.

\section{Solution of the periodic Toda lattice}
\renewcommand{\theequation}{3.\arabic{equation}}\setcounter{equation}{0}
\renewcommand{\thefigure}{3.\arabic{figure}}\setcounter{figure}{0}
Here we give a brief summary of the theory of periodic Toda lattice and
the solution formula in terms of the Riemann theta function ${\vartheta}
(x|B)$. This will be used in the next section to study the geometry of the
divisors which
are given by the zeros of $\vartheta(x|B)$ with some translates in $x$.

\subsection{Isospectral manifold and symmetric products of Riemann surfaces}
Let a pair of $\infty$-vectors $\{\phi^{(1)},\phi^{(2)}\}$  be a fundamental set of solutions for
the difference equation $L\phi=\lambda \phi$ with the periodic condition
(\ref{ab}). The $\phi^{(j)} =(\cdots, \phi^{(j)}_{-1}, \phi^{(j)}_0,
\phi^{(j)}_{1},\cdots)$ are assumed to satisfy
\begin{equation}
\label{phi12}
\left\{
\begin{array}{ll}
\phi^{(1)}_0&= 1,\\
\phi^{(1)}_{1} &=0,
\end{array}
\right. \quad {\rm and} \quad
\left\{
\begin{array}{ll}
\phi^{(2)}_0 &=0, \\
\phi^{(2)}_{1} &=1.
\end{array}
\right.
\end{equation}
(Note here that the choice of the index ``0" can be shifted by an arbitrary number $k$ (mod $N+1$).)
Then one can define a monodromy matrix $M$ by setting
\begin{equation}
\label{monodromy}
\left(
\begin{matrix}
\phi^{(1)}_{k+N} \\
\phi^{(2)}_{k+N}
\end{matrix}
\right) =M \left(
\begin{matrix}
\phi^{(1)}_k \\
\phi^{(2)}_k
\end{matrix}
\right), \quad {\rm with} \quad
M=\left(
\begin{matrix}
\phi^{(1)}_{N} &\phi^{(1)}_{N+1} \\
\phi^{(2)}_{N} &\phi^{(2)}_{N+1}
\end{matrix}
\right)~.
\end{equation}
First we have:
\begin{Lemma}
With the condition $\prod_{k=1}^N a_k=1$, we have
${\rm det}M=1$.
\end{Lemma}
\begin{Proof}
One can easily show that the Wronskian $W(n):=\phi^{(1)}_n\phi^{(2)}_{n+1}
-\phi^{(2)}_n\phi^{(1)}_{n+1}$ satisfies
\begin{equation}
\nonumber
W(n)=a_{n-1}W(n-1)=\left(\prod_{k=0}^{n-1}a_k\right)W(0).
\end{equation}
Then put $n=N$, i.e. $W(N)={\rm det}M$, and use $W(0)=1$ to complete the
proof.
\end{Proof}
We can also see from the periodic condition $\phi_{k+N}=z\phi_k$ in
(\ref{phiperiod})
that the spectral parameter $z$ is the eigenvalue of $M$, and we have,
\begin{equation}
\nonumber
\displaystyle{z+{1\over z}-{\rm tr}M=0.}
\end{equation}
Then we have
\begin{equation}
\label{traceM}
{\rm tr}M=\phi^{(1)}_N+\phi^{(2)}_{N+1}=P(\lambda),
\end{equation}
where the $N$-th polynomial $P(\lambda)$ is given by (\ref{P}).

Let us now find the explicit form of the fundamental solutions
$\phi^{(j)}$.
By a direct computation, one can easily find:
\begin{Lemma}
The $\phi^{(j)}_{n+1}$ for $n\ge 1$ are given by
\begin{equation}
\label{phi}
\phi^{(1)}_{n+1}=-a_N(-1)^{n-1}\Delta_{2,n}, \quad \phi^{(2)}_{n+1}
=(-1)^n\Delta_{1,n}~,
\end{equation}
where $\Delta_{m,n}$ are defined by (\ref{Deltamn}) and $\Delta_{2,1}:=1$.
\end{Lemma}
Then we consider the so-called auxiliary spectrum $\mu_n$'s which are
defined by the roots of the equation $\phi^{(1)}_{N+1}(\lambda)=0$, and we
denote
\begin{equation}
\label{mu}
u_g(\lambda):=(-1)^{N-1}\Delta_{2,N}(\lambda)=
\displaystyle{\prod_{k=1}^{N-1}(\lambda-\mu_k)} =0, \quad {\rm with} ~
g=N-1.
\end{equation}
Note that some of the roots are complex, and they form a pair of
complex conjugate roots $(\mu_k, {\bar \mu}_k)$.
With (\ref{mu}), the solution of the Toda lattice $a_k$'s and $b_k$'s are
expressed by
the symmetric functions of the roots $\mu_j$'s under the isospectral
conditions, $I_k=\gamma_k$ for $k=1,\cdots,N+1$. For example,
 we have from (\ref{mu})  and $I_1(L_P)=0$
\begin{equation}
 \label{b}
	 b_1=-\displaystyle{\sum_{j=2}^N b_j= -\sum_{n=1}^{N-1} \mu_n}.
\end{equation}
This implies:
\begin{Proposition}
\label{Jmap}
The map $\varphi$ from $Z_{\Bbb R}^P(\gamma)$ to the real part of
the $g$-fold symmetric product of the affine part $\mathcal R_0$,
denoted as ${\mathcal R}^{(g)}_0:={\mathcal R}^g_0/S_g$,
\begin{equation}
\nonumber
\varphi :  Z_{\Bbb R}^P(\gamma) \longrightarrow
\Big({\mathcal R}_0^{(g)}\Big)_{\Bbb R}
\end{equation}
is surjective.
Here $S_g$ is the symmetry group of order $g=N-1$.
\end{Proposition}
The formulation of the Toda lattice with the variable $\mu_k$'s is
called the Mumford system \cite{mumford:84} which parametrizes the
moduli space of the hyperelliptic Riemann surface in terms of the
triples of polynomials, $(u_g(\lambda), v_g(\lambda), w_g(\lambda))$, defined
by the factorization of the curve (\ref{y}),
\begin{equation}
\label{mumfordsystem}
y^2=u_g(\lambda)w_g(\lambda)+v_g^2(\lambda).
\end{equation}
Here $u_g(\lambda)$ is given by (\ref{mu}), $v_g(\lambda)$ is
a polynomial of degree $g-1=N-2$, and
$w_g(\lambda)$ is a monic polynomial of degree $g+2$. Note here that if all
the roots $\mu_k$'s
are distinct, the $v_g(\lambda)$ can be expressed in terms of
$\mu_k$'s with the condition $v_g(\mu_k)=y(\mu_k)$, i.e.
\begin{equation}
\label{v}
v_g(\lambda)=\displaystyle{\sum_{k=1}^{g} y(\mu_k)\prod_{j\ne k}
{\lambda -\mu_j \over \mu_k-\mu_j}.}
\end{equation}
The Mumford systems of the
triples $(u_d(\lambda), v_d(\lambda), w_d(\lambda))$ with $0< d<g$
can be also used to parametrize the divisors (see the next section).

Now we discuss the topology of the set of the roots $\mu_k$'s:
We first note that $\pi({\mathcal R}_{\Bbb R})$ with $\pi:{\mathcal R}\to
\pi({\mathcal R}),
~P=(\lambda,y)\mapsto \pi(P)=\lambda$, consists of some intervals,
\begin{equation}
\nonumber
\pi({\mathcal R}_{\Bbb R})=\displaystyle{
\bigcup_{m=0}^{N_R}
\Lambda_m, \quad 1\le N_R\le N},
\end{equation}
where $\Lambda_m=[\lambda_{2m},\lambda_{2m+1}]$ with
$\lambda_0=\lambda_{2N_R+1}=\infty$, (i.e. $[\lambda_0,\lambda_1]$
and $[\lambda_{2N_R},\lambda_{2N_R+1}]$ are connected through
$\pi(p^{\pm}_{\infty})$).
Then we first note:
\begin{Lemma}
If the root $\mu_k$ is real, then $\mu_k$ belongs to one of the intervals
$\Lambda_{m}$.
\end{Lemma}
\begin{Proof}
Since $\phi_{N+1}^{(1)}(\mu_k)$=0, we have
\begin{equation}
\nonumber
{\rm tr}M=\displaystyle{\phi_N^{(1)}+{1\over \phi_N^{(1)}}}.
\end{equation}
This implies $|{\rm tr}M|\ge 2$, and therefore $\mu_k\in
\pi({\mathcal R}_{\Bbb R})$.
\end{Proof}
Since the set of the points $P_k\in {\mathcal R}$ consists of real ones
and/or pairs of
complex conjugates,
we have the following Lemma on the topology of those sets:
\begin{Lemma}
\label{mus}
\cite{mumford:84}
The set of points $P_k\in {\mathcal R}$ with $\pi(P_k)=\mu_k,~k=1,\cdots,g$
can be classified as
\begin{itemize}
\item[a)] $\mu_k \in \Lambda_{m_k}$ for some $m_k\in\{0,1,\cdots,N_R\}$,
and the set of those points for each $k$
is isomorphic to $S^1$, or
\item[b)] ${\overline{P}}_{s+1}=P_{s+2}$, or both $\mu_{s+1}, \mu_{s+2}$
in the same
$\Lambda_{m_s}$ for some $m_s$, and the set of those points is determined by
$V=\{(P,{\overline{P}})\in {\mathcal R}^{(2)}={\mathcal R}^2/S_2\}$.
\end{itemize}
\end{Lemma}
 The later case $b)$
 is obtained by considering two points $P,{\overline{P}} \in
{\mathcal R}$, which we can move continuously on ${\mathcal R}$
until $P={\overline{P}}$ with $\pi(P)\in \Lambda_{m_s}$,
and then move the two points independently on the real loop over
$\Lambda_{m_s}$. The set of points in this case 
includes $p_{\infty}^{\pm}$. From this Lemma, we can see that the solution
corresponding to the case $a)$ except the case $\mu_k\in \Lambda_0 \cup
\Lambda_{N_R}$ is just a periodic cycle, and
the solutions to all other cases blow up.
\vskip 0.5cm

Now we can give a proof of Propositions \ref{realJac}
(see also \cite{audin:96, gross:81}):
\begin{Proof}
Let us first denote ${\mathcal R}_{\Bbb R}=C_0\sqcup\cdots\sqcup C_m$ with
$m=N_R-1$, where each $C_i$ gives a set corresponding to the case $a)$ in
Lemma \ref{mus}, that is, $P\in C_i$ implies $\pi(P)\in \Lambda_i$.
Then Lemma \ref{mus} implies that the symmetric product
${\mathcal R}^{(g)}_{\Bbb R}=({\mathcal R}^g/S_g)_{\Bbb R}$ is covered by
the sets,
$$
U(i_1,\cdots,i_j):=C_{i_1}\times \cdots \times C_{i_j}\times V^k, \quad
{\rm with}\quad j+2k=g.
$$
Then the number of disconnected sets can be computed as follows:
Let $(n_1,\cdots,n_g)$ be an increasing sequence of numbers with
$n_k\in\{0,1,\cdots,m\}$
and $m=N_R-1$, such that if $s\in\{1,\cdots,g\}$ is
the
minimum number for the first nonzero $n_k$, then $n_k$'s in the sequence
assume,
\begin{equation}
\nonumber
0=n_1=\cdots=n_{s-1}<n_s<n_{s+1}<\cdots<n_g \le m.
\end{equation}
Then we associate the number $n_k$ in the sequence with the cases
in Lemma \ref{mus} as follows:
\begin{itemize}
\item[i)] $n_k=0$ implies that either $\mu_k$ belongs to
$\Lambda_0\cup\Lambda_M$,
or it is in the case $b)$, i.e. $P_k\in C_0$ or $(P_k, P_l)\in V$ for some
$P_l={\overline {P}}_k$,
\item[ii)] $n_k\ne 0$ implies that $\mu_k$ belongs to a bounded interval
$\Lambda_{n_k}=[\lambda_{2n_k},\lambda_{2n_k+1}]$.
\end{itemize}
Then counting the number of sequences $(n_1,\cdots,n_g)$ with
$n_k\in\{0,1,\cdots,m\}$
for each case of ${\mathcal R}_{\Bbb R}$ having $N_R=m+1$ connected components,
we obtain the number of connected components of
${\mathcal R}^{(g)}_{\Bbb R}$ as
\begin{equation}
\nonumber
n\left({\mathcal R}^{(g)}_{\Bbb R}\right)=\displaystyle{
\sum_{j=0}^m \binom{m}{j}=2^m}.
\end{equation}

On the other hand, writing the anti-holomorphic
involution in a matrix form on the lattice 
given by the period matrix $\Omega=(I,B)$ in
(\ref{periodicmatrix}), one can show that each connected component
of $({\Bbb C}^g/\Gamma)_{\Bbb R}\cong{\rm Jac}({\mathcal R})_{\Bbb R}$
is diffeomorphic to the real torus ${\Bbb R}^g/{\Bbb Z}^g={\Bbb T}^g$.

This then completes the proof by noticing $n({\rm Jac}_{\Bbb R})=
n({\mathcal R}^{(g)}_{\Bbb R})$ (see also below).
\end{Proof}
Note here that the number $\binom{m}{j}$ represents the total number of the
sequences having the form which is identified as the set $U(i_1,\cdots,i_j)$
as
\begin{equation}
\label{symmcomponent}
(\overbrace{0,\cdots,0}^{g-j=s-1},\overbrace{n_{s},\cdots,n_g}^{j}) \equiv
\left\{
\begin{array}{lll}
V^k\times C_{n_{s}}\times\cdots\times C_{n_g} & {\rm if} & s=2k+1 \\
 V^k\times C_0\times C_{n_{s}}\times\cdots\times C_{n_g} & {\rm if} & s=2k+2,
\end{array}
\right.
\end{equation}
with $n_{s}\ne 0$. Thus each connected component can be characterized by
the number of components $C_{n_i}$ with nonzero $n_i$.

\subsection{Toda flow on the Jacobian}
Let us now find the Toda flow on the Riemann surface which is
determined by the image of the dynamics of $\mu_n$'s
under the $g$-hold Abel-Jacobi
map $\upsilon_g:{\mathcal R}^{(g)}\to {\rm Jac}({\mathcal R})_{\Bbb R}$.
First we note that the time evolution of $\phi^{(1)}_n$ satisfies
\begin{equation}
\nonumber
{d \phi^{(1)}_n\over dt}
+a_{n-1}\phi^{(1)}_{n-1}=a_{N-1}\phi^{(1)}_{-1}\phi^{(1)}_n+a_N\phi^{(2)}_n.
\end{equation}
which can be shown by computing the $n$-th element of the equation,
${d \phi^{(1)}\over dt}-A\phi^{(1)}
=c_1\phi^{(1)}+c_2\phi^{(2)}$.
Then the dynamics of $\mu_n$ can be obtained by evaluating the equation
with (\ref{phi})
at $\lambda=\mu_j$ and $n=N+1$, i.e.
\begin{equation}
\nonumber
\displaystyle{{d\over dt}\left(-a_N\prod_{k=1}^{N-1}
(\lambda-\mu_k)\right)\Big|_{\lambda=\mu_j}+a_N\phi^{(1)}_N(\mu_j)
=a_N\phi^{(2)}_{N+1}(\mu_j).}
\end{equation}
Noting $\phi^{(2)}_{N+1}=1/\phi^{(1)}_N$ and using (\ref{traceM}),
we obtain
\begin{equation}
\label{mumfordequation}
\displaystyle{{1\over y(\mu_j)}{d \mu_j\over dt}=\pm
{1\over \prod_{k\ne j}(\mu_j-\mu_k)}},
\end{equation}
where $y(\lambda)=\sqrt{P(\lambda)^2-4}$. Using the Lagrange interpolation
formula,
\begin{equation}
\nonumber
\displaystyle{\sum_{j=1}^{N-1}{\mu_j^n \over \prod_{k\ne j}(\mu_j-\mu_k)}=}
\left\{
\begin{array}{ll}
0 \quad {\rm if} ~ n<N-2,\\
1  \quad {\rm if} ~ n=N-2. \\
\end{array}
\right.
\end{equation}
we obtain, after the integration,
\begin{equation}
\label{muJacobi}
\displaystyle{\sum_{j=1}^{N-1}\int_{\mu_*}^{\mu_j}{\lambda^n d\lambda \over
y(\lambda)}}=\left\{
\begin{matrix}
\delta_n  & n< N-2,\\
\pm t+\delta_{g-1} & n=N-2. \\
\end{matrix}
\right.
\end{equation}
These equations can be also written in terms of the holomorphic differentials
$\omega_k$ normalized as (\ref{normalization}), and we now have the well-known
result (see for example \cite{toda:81}):
\begin{Proposition}
\label{jacobiansol}
 The periodic Toda flow can  be expressed as a linear flow on the Jacobian
${\rm Jac}({\mathcal R})\cong {\Bbb C}^g/{\Gamma}$
associated with the Riemann surface $\mathcal R$ of genus $g=N-1$, i.e.
\begin{equation}
\nonumber
\displaystyle{\sum_{j=1}^g\int_{\mu_*}^{\mu_j}\omega_k=
\pm C_{k,g-1}t+\delta'_k, \quad ({\rm mod~} \Gamma) \quad k=1,\cdots,g},
\end{equation}
where $\delta'$ is a constant, and
the $g\times g$ matrix $(C_{k,n})$ is determined by (\ref{normalization})
with
\begin{equation}
\label{omegadiff}
\omega_k=\displaystyle{\sum_{n=0}^{g-1}C_{k,n}{\lambda^n d\lambda \over y}}.
\end{equation}
\end{Proposition}
With Propositions \ref{Jmap} and \ref{jacobiansol}, we have constructed the
maps,
\begin{equation}
\label{Jumap}
\begin{matrix}
Z_{\Bbb R}^P(\gamma) & \overset{\varphi}{\longrightarrow}
& {\mathcal R}_{\Bbb R}^{(g)} &\overset{\upsilon_g}
{\longrightarrow}& Jac({\mathcal R})_{\Bbb R} \\
(b_1,\cdots,b_N) &\longmapsto & [\mu_1,\cdots,\mu_g] &\longmapsto &
\left[\sum_{j=1}^g
\int_{\mu_*}^{\mu_j}\omega\right]\\
\end{matrix}
\end{equation}
where $\omega=(\omega_1,\cdots,\omega_g)$, and the $g$-fold Abel-Jacobi map
$\upsilon_g$ is surjective \cite{mumford:84}.

Since the preimage of each connected component in ${\rm Jac}({\mathcal R})_{\Bbb R}$ 
is given by (\ref{symmcomponent}) in ${\mathcal R}^{(g)}_{\Bbb R}$, we denote
\begin{equation}
\label{connectedcomponent}
 \left(
{\Bbb T}^g\right)_j\cong (0,\cdots,0,n_{g-j+1},\cdots,n_g), \quad {\rm
for}\quad 0\le j \le m \le g,
\end{equation}
so that we have
\begin{equation}
\label{connectedcomponentJ}
{\rm Jac}({\mathcal R})_{\Bbb R} \cong \bigsqcup_{j=0}^m {\mathcal M}_j,
\quad {\mathcal M}_j=\overbrace{({\Bbb T}^g)_j\sqcup\cdots\sqcup({\Bbb
T}^g)_j}^{\binom{m}{j}}.
\end{equation}
\begin{Remark}
Since $({\Bbb T}^g)_j$ is a real part of a complex abelian group
${\rm Jac}({\mathcal R})\cong {\Bbb C}^g/\Gamma$, $({\Bbb T}^g)_j$
is a real abelian group diffeomorphic to a $g$-dimensional real torus
${\Bbb R}^g/{\Bbb Z}^g$. We will later introduce $({\Bbb T}^k)_j$
 with
\[
({\Bbb T}^k)_j\cong (\overbrace{0,\cdots,0,n_{g-j+1},\cdots,n_g}^k),
\quad {\rm for} \quad j\le k< g.
\]
However this may not be a $k$-dimensional real torus except
the case with $k=j$. In this paper, we will not discuss
the details of $({\Bbb T}^k)_j$, but its topology can be determined
by the set $V=\{(P,{\bar P})\in {\mathcal R}^{(2)}\}$
(see \cite{vanhaecke:98}). The details will be given elsewhere.
\end{Remark}

\begin{Remark}
Equation (\ref{mumfordequation}) can be expressed as an integrable system
of the Mumford system (\ref{mumfordsystem}) \cite{vanhaecke:98}:
\begin{equation}
\label{mumfordintegrable}
\displaystyle{{dM_g(\lambda) \over dt}=[ M_g(\lambda), B_g(\lambda)],}
\end{equation}
where the matrices $M_g(\lambda)$ and $B_g(\lambda)$ are given by
the triples $(u_g, v_g, w_g)$ of the Mumford system,
\begin{equation}
\nonumber
M_g(\lambda)=
\left(
\begin{matrix}
v_g(\lambda) & u_g(\lambda)\\
w_g(\lambda) & -v_g(\lambda)
\end{matrix}
\right), \quad
B_g(\lambda)={1\over 2}
\left(
\begin{matrix}
0 & 1 \\
b_g(\lambda) & 0
\end{matrix}
\right).
\end{equation}
Here $b_g(\lambda)$ is a monic polynomial of degree $g+2$ which is uniquely
determined by
the consistency of the system (\ref{mumfordintegrable}).
The equation can be generalized to describe the dynamics
on the devisor (see below).
\end{Remark}

\subsection{Riemann theta function and $\Theta$ divisor}
Here we give the solution of the periodic Toda lattice in terms of
the Riemann theta function. We also discuss the geometry of the set
of blow-up points determined by the $\Theta$ divisor, the zeros
of the Riemann theta function.

Let us first define the Siegel upper half plane ${\mathfrak S}_g$
as the set of symmetric $g\times g$ complex matrices with positive definite
imaginary part, i.e.
\begin{equation}
\nonumber
{\mathfrak S}_g:=\left\{ B\in {\mathfrak M}(g,{\Bbb C}): B^t=B, ~ {\rm
Im}~B>0\right\}.
\end{equation}
Then on the product ${\Bbb C}^g\times {\mathfrak S}_g$ we can define
 the Riemann theta function
\begin{equation}
\nonumber
\vartheta (z, B)=\displaystyle{\sum_{m\in {\Bbb Z}^g}
{\bf e}\left(m^tz+{1\over 2}m^tBm\right).}
\end{equation}
where ${\bf e}(x)=e^{2\pi\sqrt{-1}x}$. Since ${\rm Im}~B>0$, the above
series converges uniformly on compact sets on ${\Bbb C}^g\times
{\mathfrak S}_g$ to a holomorphic function.  It is easy to see that the
theta function satisfies the quasi-periodic relations,
\begin{equation}
\nonumber
\left\{
\begin{array}{lll}
\vartheta (z+E_k, B) & = & \vartheta (z, B) \\
\vartheta (z+B_k, B) & = & {\bf e}(-z_k-B_{k,k})\vartheta (z, B)\\
\end{array}
\right.
\end{equation}
where $E_1, \cdots, E_g,B_1, \cdots, B_g$ are the $2g$ column vectors in the
period matrix $\Omega=(I~B)$. It follows from those equations that for each
fixed $B\in {\mathfrak S}_g$ the divisor of the Riemann theta
function
is invariant under translations by elements of the lattice $\Gamma$
generated by
$E_1,\cdots, B_g$ and it induces a divisor $\Theta_0$ on ${\rm Jac}({\mathcal
R})$,
which is called the theta divisor,
\begin{equation}
\nonumber
\Theta_0:=\left\{[z]\in {\rm Jac}({\mathcal R}):\vartheta (z, B)=0\right\}.
\end{equation}

Let us consider a function $f:{\mathcal R}\to {\Bbb C}$ defined by
\begin{equation}
\nonumber
f(P):= \vartheta ({\tilde \upsilon}(P)-c, B)
\end{equation}
where $c\in {\Bbb C}^g$ is an arbitrary constant vector,
and $\tilde \upsilon$ is a lift of the Abel-Jacobi map $\upsilon$, i.e.
\begin{equation}
\nonumber
{\tilde \upsilon}: {\mathcal R}' \longrightarrow  {\Bbb C}^g;
\quad P  \longmapsto \left(\int_{P_0}^P \omega\right)
 \end{equation}
  with ${\mathcal R}'$
the $4g$-sided polygon obtained by cutting $\mathcal R$ along the symplectic
basis, $\{(\alpha_k,\beta_k):k=1,\cdots,g\}$,
and $\omega=(\omega_1,\cdots,\omega_g)$ the holomorphic differential.
Then the following Lemma is standard (e.g. see \cite{farkas:80} for
the proof):
\begin{Lemma}
The followings are satisfied for the zeros of the function $f(P)$:
\begin{itemize}
\item[a)]  $f(P)$ takes exactly $g$ zeros on ${\mathcal R}$ including their
multiplicities.
\item[b)] If $P_1,\cdots,P_g$ are the zeros of $f(P)$,
then
\begin{equation}
\nonumber
\displaystyle{\sum_{j=1}^gu(P_j)+[\kappa]=[c].}
\end{equation}
where $\kappa$ is the Riemann constant.
\end{itemize}
\end{Lemma}

This implies
\begin{Proposition}
\label{gzeros}
Let $Q_1,\cdots,Q_g$ be a set of arbitrary points on ${\mathcal R}$. Then
the function
$\vartheta\left({\tilde \upsilon}(P)-\sum_{j=1}^g{\tilde \upsilon}(Q_j)-\kappa\right)$
has $g$ zeros precisely at $P=Q_1,\cdots,Q_g$.
\end{Proposition}
Geometrically, Proposition \ref{gzeros} provides the well-known
fact on the theta divisor $\Theta_0$,
\begin{Corollary}
\label{W}
The theta divisor $\Theta_0$ of ${\rm Jac}({\mathcal R})$ satisfies
\begin{equation}
\Theta_0 =W^{g-1}+[\kappa]
\end{equation}
where $W^{g-1}$ is the image of the $r$-fold Abel-Jacobi map $\upsilon_r$ with
$r=g-1$, i.e.
\begin{equation}
\begin{matrix}
\upsilon_{r}:&{\mathcal R}^{(r)} &\longrightarrow &{\rm Jac}({\mathcal R})\\
  &  [Q_1,\cdots,Q_{r}] & \longmapsto & \left[\sum_{j=1}^{r}\int_{Q_0}^{Q_j}
  \omega\right]\\
 \end{matrix}
\end{equation}
 \end{Corollary}

\begin{Example}
$g=2$ Riemann surface and the divisor $\Theta_0$: 
In the case of $g=2$ compact Riemann surface $\mathcal R$, the image $W^1$
is isomorphic to
the Riemann surface, and thus the Jacobian ${\rm Jac}({\mathcal R})$ contains
a $g=2$ Riemann surface as a closed submanifold, $\Theta_0\cong {\mathcal R}$.
Thus we have
${\rm Jac}({\mathcal R})_{\Bbb R}\cap \Theta_0\cong {\mathcal R}_{\Bbb R}$.

\end{Example}

Now we can find the solution of the periodic Toda lattice by solving
the Jacobi inversion problem for $\mu_j$'s in (\ref{jacobiansol}).
Then the solution $\mu_j(t)=\pi(P_j)$
 is given by the zero of the theta function,
\begin{equation}
\nonumber
f(P)=\vartheta({\tilde \upsilon}(P)-c, B)=0, \quad {\rm with} ~~ c=\pm
C_{g-1}t+\kappa',
\end{equation}
where $\kappa'=\kappa+ \delta'$, and $C_{g-1}=(C_{1,g-1},\cdots,C_{g,g-1})$
is the column vector of the matrix $(C_{k,n})$ in (\ref{omegadiff}).
It follows from (\ref{b}) that
 we need just the sum of $\mu_n$'s for the solution $b_k(t)$. The sum
 $\sum_{j=1}^g\mu_j$ can be obtained by the integral, that is,
\begin{equation}
\nonumber
\displaystyle{{1\over 2\pi\sqrt{-1}}\int_{\partial {\mathcal R}'}
\pi (P) d \log f(P)=\sum_{j=1}^g\mu_j +\pi(p_{\infty}^+)+\pi
(p_{\infty}^-),}
\end{equation}
where $\pi (p_{\infty}^{\pm})$ are the contributions from the infinity.
Finally we obtain (see \cite{toda:81} for the details of the proof):
\begin{Proposition}
The solution $(a_1,\cdots,a_N,b_1,\cdots,b_N)$ of the Toda lattice is given by
\begin{equation}
\label{solutionabk}
\left\{
\begin{array}{ll}
a_k(t) &=\displaystyle{{\tau_{k+1}(t)\tau_{k-1}(t)\over (\tau_k(t))^2}}\\
b_{k}(t)&=\displaystyle{{d \over dt} \log{\tau_k(t) \over \tau_{k-1}(t)}}
\end{array}
\right.
\end{equation}
where $\tau_k(t)={\vartheta(c't+ck+\delta_0, B)}$ with $c'=\pm C_{g-1},
c={\tilde \upsilon}(p_{\infty}^-)-{\tilde \upsilon}(p_{\infty}^+)$
and $\delta_0={\tilde \upsilon}(p_{\infty}^+)-\kappa-\delta'$.
\end{Proposition}
Now it is clear that if a $\tau_k(t)$ vanishes for some $k$ and $t$,
the functions $a_k(t)$ and $b_{k\pm 1}(t)$ blow up. Thus the devisor
$\Theta$ is given by the sum of the zero sets of $\tau$-functions,
which are the theta devisor $\Theta_0$ and
its translates $\Theta_k$, i.e.
\begin{equation}
\label{sumdivisor}
\Theta =\bigcup_{k=0}^g \Theta_k, \quad {\rm with}\quad \Theta_k=\Theta_0+
k({\tilde \upsilon}(p_{\infty}^-)-{\tilde \upsilon}(p_{\infty}^+)).
\end{equation}
Thus
the isospectral set $Z_{\Bbb R}^P(\gamma)$ can be compactified by
gluing on the $\Theta$ divisors. One can also observe that the multiplicity of
the zero of $\tau_k(t)$ gives the intersection multiplicity of the
divisors. The complex version of the geometry
of the divisors has been worked out in \cite{adler:91}.
In this paper, we are concerned with the real version of this study, and
our goal is to give a Lie theoretic study on the gluing structure in terms
of a representation of the affine Weyl group $W$.

\begin{Remark}
With the solution form (\ref{solutionabk}), one can easily show that
the $tau$-functions $\tau_k(t)$ satisfy the Hirota bilinear form,
\begin{equation}
\label{bilinear}
\tau_k\tau_k''-(\tau_k' )^2=\tau_{k+1}\tau_{k-1},
\end{equation}
where $\tau_k'=d\tau_k/dt$ and $\tau_k''=d^2\tau_k/dt^2$.
This formula will be useful for the study of degeneracy of the zeros,
which characterizes the intersection of the divisors.
\end{Remark}

\section{Conjectures on the topology of ${\rm Jac}(\mathcal R)_{\Bbb R}$}
\renewcommand{\theequation}{4.\arabic{equation}}\setcounter{equation}{0}
\renewcommand{\thefigure}{4.\arabic{figure}}\setcounter{figure}{0}
Here we provide several conjectures on the geometry
of the divisor on ${\rm Jac}(\mathcal R)_{\Bbb R}$ and
the topology of the affine part of
${\rm Jac}({\mathcal R})_{\Bbb R}$ .

\subsection{Geometry of the divisor $\Theta$}
Recall that $\Theta_k$ is the translate of the theta divisor $\Theta_0$
by $k(Q-P)$ with $Q={\upsilon}(p^-_{\infty}), P={\upsilon}(p^+_{\infty})$,
so that $\Theta_k\cong \Theta_0\cong W^{g-1}$.
Then following
\cite{adler:91}, we define the sets $\Theta_J$
for each subset $J\subsetneqq \{0,1,\cdots,g\}$ as the intersections with
$\Theta_k$,
\begin{equation}
\Theta_{J}:=\bigcap_{k\in J} \Theta_k, \quad {\rm and} \quad
{\rm dim}_{\Bbb C}\Theta_J=g-|J|.
\label{thetaJ}
\end{equation}
Note here that
$\Theta_J\supset\Theta_{J'}$ if $J\subset J'$, and
$\Theta_J\cong {\mathcal R}$ if $|J|=g-1$. We then have
a stratification of the Jacobian,
\[
{\rm Jac}({\mathcal R})\supset \Theta^{(1)}\supset\Theta^{(2)}\supset
\cdots\supset\Theta^{(g)}, \quad {\rm with}\quad
\Theta^{(k)}:=\bigcup_{|J|=k}\Theta_J.
\]
We also have a cell decomposition of the devisor $\Theta$
in terms of those sets,
\begin{equation}
\label{thetadecomposition}
\Theta =\bigsqcup_{J\subsetneqq \{0,1,\cdots,g\}}D_J,
\quad {\rm with}\quad D_J:=\Theta_J\bigcap_{k\notin J}\Theta_{k}^c,
\end{equation}
where $\Theta^c_k$ is the complement set of $\Theta_k$ in ${\rm Jac}({\mathcal R})$.
In order to compute the intersections ${\rm Jac}({\mathcal R})_{\Bbb R}
\cap \Theta_J$, we first state:
\begin{Lemma}
\label{intersectionDT}
Let $({\Bbb T}^g)_j$ be a $g$-dimensional torus defined by
(\ref{connectedcomponent}). Then we have
\[
({\Bbb T}^g)_j\cap\Theta_J=
\left\{
\begin{matrix}
({\Bbb T}^{g-|J|})_j & {\rm if} & |J|\le g-j, \\
\emptyset & {\rm if} & |J|> g-j.
\end{matrix}
\right.
\]
\end{Lemma}
\begin{Proof}
Let us first consider the case of $J=\{k\}$ for $k=0,1,\cdots,g$.
Since $\Theta_k$ is isomorphic to the image of the $(g-1)$-hold
symmetric product of the Riemann surfaces, $\Theta_k\cong W^{g-1}$
(Proposition \ref{W}),
$\Theta_k\cap ({\Bbb T}^g)_j$ consists of the elements
denoted by $(\overbrace{0,\cdots,0}^{g-j-1},n_{g-j+1},\cdots,n_g)$,
that is, one of the first $(g-j)$ elements in $({\Bbb T}^g)_j$
is fixed at a point of infinity. This then implies the Lemma.
In the case of $|J|>1$, it is not hard to see that a similar
argument can lead to the proof.
\end{Proof}
\begin{Remark}
Lemma \ref{intersectionDT} is a consequence of the Poincar\'e's formula,
\[
\Theta_{J}\cong |J|!~ W^{g-|J|} \quad ({\rm homologous})
\]
and the variety $({\Bbb T}^{g-|J|})_j$ is singular in general.
It seems however that the $({\Bbb T}^{g-|J|})_j$ is smooth
if $J$ consists of a consecutive numbers $J=\{s+1,\cdots,s+|J|\}, ({\rm mod} N)$
(see Theorem 5. and Corollary 1. in \cite{adler:91}).
\end{Remark}
Since the divisor $\Theta$ is the sum of the zeros of tau-functions,
$\tau_k(t)=\theta (k(Q-P)+R(t))$, the cell $D_J$ is also defined by
\[
R(t_J)\in D_J\cap{\rm Jac}({\mathcal R})_{\Bbb R}
\overset{\rm def}{\iff} \tau_k(t_J)=0, ~{\rm iff} ~k\in J.
\]
On $D_J$, one can determine the multiplicity $m_k$ of the
zero of $\tau_k(t)$ with $k\in J$, which describes the intersection
geometry of the divisors $\Theta_k$ with $k\in J$.
\begin{Lemma}
\label{multiplicity}
Let $J$ be a set of consecutive numbers given by $\{i+1,\cdots,i+s\}$
(mod $N$).
Then $\tau_k(t)$ has the following form near its zero $t=t_J$,
\begin{equation}
\label{taukzero}
\tau_{i+k}(t)\simeq (t-t_J)^{m_{k}}+\cdots, \quad {\rm with}\quad m_k=k(s+1-k),
~ 1\le k\le s.
\end{equation}
\end{Lemma}
\begin{Proof}
Substituting (\ref{taukzero}) into (\ref{bilinear}), and using
$\tau_i(t_J)\ne 0$, we have $m_k=k(m_1+1-k)$.
Then $\tau_{i+s+1}(t_J)\ne 0$ implies $m_1=s$.
\end{Proof}
As was shown in \cite{adler:93}, the divisor $D_J$ can be parametrized
by the so-called {\it limit matrix}, say $L_J$, defined by
\begin{equation}
\label{limitmatrix}
L_J:= \lim_{t\to t_J} n^{-1}_J(t)L(t)n_J(t), \quad {\rm for~some}\quad
n_J(t)\in N_-,
\end{equation}
where $L_J$ has $2N-|J|$ variables, and $N_-$ is the set of lower
triangular matrices with $1$'s on the diagonals. The matrix $n_J$ is determined uniquely by
the structure of the blow-up given in Lemma \ref{multiplicity}. For the set $J=\{i_1,
\cdots,i_n\}$, the limit matrix $L_{J}$
can be constructed from a chain of matrices starting from the
original matrix $L_P$,
\[
L_P\to L_{J_1}\to L_{J_2}\to \cdots \to L_{J}~,
\]
where $J_k=\{i_1,\cdots,i_k\}$ with $k< n=|J|$ and $J_n=J$. The matrices $n_{J_k}\in N_-$
can be found easily in terms of $L_{J_k}$, and then we have
the limit $n^{-1}_{J_k}L_{J_k}n_{J_k} \to L_{J_{k+1}}$. Then the $L_{J_k}$ parametrizes
the $D_{J_k}$ devisor of ${\rm codim}_{\Bbb C}D_{J_k}=k$
(we will report the details
elsewhere).

Now with the limit matrix $L_J$, one can also define an integrable
system on $D_J$ in terms of the Mumford triples $(u_d(\lambda),
v_d(\lambda),w_d(\lambda))$ with $d={\rm dim}_{\Bbb C} D_J=g-|J|$
\cite{mumford:84, vanhaecke:98}. Here the triples satisfy
\[
y^2=P_J(\lambda)^2-4=u_dw_d+v^2_d, \quad {\rm det}(L_J-\lambda I)=
-(z+z^{-1}-P_J(\lambda)),
\]
where $u_d(\lambda)$ is a monic polynomial
of degree $d$, $v_d(\lambda)$ is a polynomial of degree $d-1$
with $v_d(\mu_j)=y(\mu_j)$ whenever $u_d(\mu_j)=0$, and
$w_d(\lambda)$ is a monic polynomial of degree $2g+2-d$.
The zeros of $u_d(\lambda)$ parametrize the variety $W^d\subset {\rm Jac}
({\mathcal R})$, the image of the symmetric products ${\mathcal R}^{(d)}$ under 
the $k$-hold Abel-Jacobi map (see \cite{mumford:84}
for this statement in terms of the divisor ${\rm Div}^{+,d}_0({\mathcal R})$).
In particular, if $J$ is given by a set of $(g-d)$ consective numbers,
say $J=\{1,\cdots,g-d\}$, then $u_d(\lambda)$ can be expressed as
\[
\begin{array}{ll}
u_d(\lambda)&=(-1)^d\Delta_{N-d+1,N}(\lambda)=\lambda^d-\left(\sum_{i=N-d+1}^N b_i
\right)\lambda^{d-1}+\cdots\\
&=\prod_{i=1}^d (\lambda-\mu_i),
\end{array}
\]
where $\delta_{m,n}(\lambda)$ is given by (\ref{Deltamn}) (see the Example
\ref{exampleg=3} below).
The integrable system on $D_J$ is then given by the same form as
(\ref{mumfordintegrable}) which gives (\ref{mumfordequation})
with the index $j$ taking $1\le j\le d=g-|J|$.
This system can be used to show the smoothness of $\Theta_J$ for this type of $J$.
Note that
the Jacobi inversion problem for such system is not complete, and
the solution manifold cannot be completed into Abelian variety
(in this sense the system is {\it not} algebraically integrable)
\cite{abenda:00}.
We will study the details of the solutions of those integrable systems
elsewhere.
\begin{Example}
\label{exampleg=3}
$D_J\subset {\rm Jac}({\mathcal R})$ of the Riemann surface of $g=3$:

\begin{itemize}
\item[i)] $J=\{1,2\}$:
The limit matrix $L_{\{1,2\}}$ is given by
\[
L_{\{1,2\}}=\left(
\begin{matrix}
0 &1 &0 & 0 \\
0 & 0& 1& 0\\
\xi_3 &-\xi_2 &\xi_1& 1-\zeta_1z^{-1}\\
z-\zeta_2 &0 &0 & \mu_1
\end{matrix}
\right),
\]
where the $3\times 3$ submatrix at the top left corner is the companion
matrix of the corresponding submatrix, i.e. $\xi_1=b_1+b_2+b_3, \xi_2=
b_1b_2+b_2b_3+b_3b_1-a_1-a_2, \xi_3=b_1b_2b_3-a_1b_3-a_2b_1$, and
$\zeta_1=-a_1a_3, \zeta_2=-a_2a_4, \mu_1=b_4$. The spectral curve associated
with $L_{\{1,2\}}$ is given by
\[
{\rm det}(L_{\{1,2\}}-\lambda I)=-(z+I_5z^{-1}-P_{\{1,2\}}(\lambda)),
\quad P_{\{1,2\}}(\lambda)=\lambda^4+\sum_{k=1}^4 (-1)^k I_k\lambda^{4-k}.
\]
Here the invariants $I_1, \cdots, I_4$ and $I_5$ are given by
\[
\left\{
\begin{matrix}
I_1=\xi_1+\mu_1=0, & I_2=\xi_1\mu_1+\xi_2, & I_3=\xi_2\mu_1+\xi_3, \\
I_4=\xi_3\mu_1+\zeta_1+\zeta_2, & I_5=\zeta_1\zeta_2=1, &
\end{matrix}
\right.
\]
The polynomials $u_1(\lambda), v_1(\lambda)$ of the Mumford triples
are just $u_1(\lambda)=\lambda-\mu_1, v_1(\lambda)=\pm y(\mu_1)$.
Then the isospectral variety in ${\rm Jac}({\mathcal R})_{\Bbb R}$
with the constraints $\tau_1=\tau_2=0$ is given by
\[
\begin{array}{ll}
D_{\{1,2\}}\cap{\rm Jac}({\mathcal R})_{\Bbb R}
& =\left\{(\xi_1,\xi_2,\xi_3,\zeta_1,\zeta_2,\mu_1)\in {\Bbb R}^6 ~:~ I_k=\gamma_k\in
{\Bbb R}, ~
k=1,\cdots,5\right\}\\
& \cong\left\{(\zeta_1,\mu_1)\in{\Bbb R}^2~:~ \zeta_1+\zeta_1^{-1}-P_{\{1,2\}}(\mu_1)=0\right\}\\
& \cong {\mathcal R}_{\Bbb R}\cap {\mathcal R}_0.
\end{array}
\]
The integrable system on this variety, $du_1(\lambda)/dt=v_1(\lambda)$,
is given by
\[
\displaystyle{{d\mu_1\over dt}=\pm y(\mu_1)=\pm\sqrt{P_{\{1,2\}}(\mu_1)^2-4}}.
\]
which describes the dynamics on the curve ${\mathcal R}_{\Bbb R}$.
\item[ii)] $J=\{1,3\}$: The limit matrix in this case is given by
\[
L_{\{1,3\}}=\left(
\begin{matrix}
0 &1 &0 & 0 \\
-\xi_2 & \xi_1& 1-\zeta_1z^{-1}& 0\\
0 &0 &0 & 1\\
z-\zeta_2 &0 &-\eta_2 & \eta_1
\end{matrix}
\right),
\]
where $\xi_1=b_1+b_2, \xi_2=b_1b_2-a_1, \eta_1=b_3+b_4, \eta_2=b_3b_4-a_3, \zeta_1=a_4b_1b_2$
and $\zeta_2=a_2b_1b_3$. The invariants are obtained as
\[
\left\{
\begin{matrix}
I_1=\xi_1+\eta_1=0, & I_2=\xi_1\eta_1+\xi_2+\eta_2, & I_3=\xi_1\eta_2+\xi_2\eta_1, \\
I_4=\xi_2\eta_2+\zeta_1+\zeta_2, & I_5=\zeta_1\zeta_2=1, & 
\end{matrix}
\right.
\]
 from which we have the isospectral variety with constraints $\tau_1=\tau_3=0$,
\[
\begin{array}{ll}
D_{\{1,3\}}\cap{\rm Jac}({\mathcal R})_{\Bbb R}
& =\left\{(\xi_1,\xi_2,\eta_1,\eta_2,\zeta_1,\zeta_2)\in {\Bbb R}^6 ~:~ I_k=\gamma_k\in
{\Bbb R}, ~
k=1,\cdots,5\right\}\\
& \cong\left\{(\zeta_1,\xi_1)\in{\Bbb R}^2~:~ \zeta_1+\zeta_1^{-1}+Q(\xi_1)=0\right\},
\end{array}
\]
where $Q(\xi_1)$ is a rational function of $\xi_1$ given by
\[
Q(\xi_1)=\displaystyle{{1\over 4}(\xi_1^2+I_2)^2-{I_3^2\over 4\xi_1^2}-I_4}.
\]
Then the corresponding divisor $\Theta_{\{1,3\}}$ in ${\rm Jac}({\mathcal R})_{\Bbb R}$ contains a 1-dimensional disconnected singular variety consisting
of two circles transversally intersecting at two distinct points of infinity.
Thus the divisors $\Theta_J$ have quite different topological structure depending on the set $J$, and
the difference may be characterized by the sub-Dynkin diagram
associated with $\tau_k=0$ for $k\in J$ (see \cite{adler:91}, and
we will report the details elsewhere).
\end{itemize} 
\end{Example}

\subsection{Affine part of the Jacobian}
Let us first state the following Conjecture on the structure of
the affine parts of ${\rm Jac}({\mathcal R})_{\Bbb R}$, which is
isomorphic to $Z_{\Bbb R}^P(\gamma)$ (see (\ref{g1Jac})
for the case of $g=1$);
\begin{Conjecture}
\label{affineJac}
The affine part of ${\rm Jac}({\mathcal R})_{\Bbb R}$ has a decomposition,
\begin{equation}
\nonumber
{\rm Jac}({\mathcal R})_{\Bbb R}\setminus{\Theta }\cong
\displaystyle{\bigsqcup_{j=0}^m
{\mathcal M}_j^{\circ} , \quad m=N_R-1\le g,}
\end{equation}
with ${\mathcal M}_j^{\circ}$ consisting of $\binom{m}{j}$ disconnected
components $({\Bbb T}^g)_j\setminus{\Theta}$
in which each torus $({\Bbb T}^g)_j$ is divided into $g+1$
 cylinders, $ {\Bbb R}^{g-j}\times {\Bbb T}^{j}$, through the divisors
 $\Theta_k$ in (\ref{sumdivisor}),
\begin{equation}
\nonumber
\left({\Bbb T}^g\right)_j\setminus \Theta=\left({\Bbb R}^{g-j}\times
{\Bbb T}^{j}\right)
\sqcup\cdots\sqcup \left({\Bbb R}^{g-j}\times {\Bbb T}^{j}\right) .
\end{equation}
\end{Conjecture}
This is motivated as follows:
Recall that $({\Bbb T}^g)_j$ is labeled by $(\overbrace{0,\cdots,0,n_{g-j+1},\cdots, n_g}^{g})$
where each element with the label $0$ contains the divisors corresponding to
$p_{\infty}^{\pm}\in {\mathcal R}$. Then removing those points through
$(g+1)$ divisor $\Theta_k$, $\Theta=\cup_{k=o}^g \Theta_k$, we expect
to have the $(g+1)$ disconneted cylinders. To prove the above Conjecture,
one needs to determine the structure of the intersections of the divisors on $({\Bbb T}^g)_j$.

We then consider the gluing pattern of those disconnected cylinders
in $({\Bbb T}^g)_j\setminus\Theta$ which are marked by the set of signs
$(\epsilon_1,\cdots,\epsilon_N)$ with $\epsilon_k={\rm sign}(a_k)$:
Let $\Sigma$ be the set of periodic sequences of signs,
\begin{equation}
\nonumber
\Sigma(N):=\left\{(\cdots,\sigma_{k-1},\sigma_k,\sigma_{k+1},\cdots):
\sigma_k\in \{\pm\},~ \sigma_{k+N}=\sigma_k,~ \forall k\in {\Bbb Z}\right\}.
\end{equation}
We also denote by $\sigma^0 \in \Sigma(N)$ the special element consisting of all $+$ signs.
We then mark each cylinder ${\Bbb R}^{g-j}\times {\Bbb T}^{j}$ with
a periodic sequence of signs, $\sigma$, and we relate the sign
$\epsilon_k={\rm sign}(a_k)$ with
$\epsilon_k=\sigma_k\sigma_{k+1}$.

We now define a permutation $s_k$ on $\Sigma(N)$,
\begin{equation}
\nonumber
\begin{matrix}
s_k: & \Sigma (N)&\longrightarrow &\Sigma (N) \\
 &  (\cdots, \sigma_k,\sigma_{k+1},\cdots) &\longmapsto & (\cdots,\sigma_{k+1},
 \sigma_k,\cdots)\\
 \end{matrix}
 \end{equation}
where $s_k$ also acts on the periodic ones $\sigma_{k+mN}$ and
$\sigma_{k+1+mN}$
for all $m\in {\Bbb Z}$, that is, $k \in {\Bbb Z}/N{\Bbb Z}$.
The set of all permutations forms an affine Weyl group $W$
associated with $A_{N-1}^{(1)}$, the affine Kac-Moody algebra of ${\mathfrak sl}(N)$,
\begin{equation}
\nonumber
{W}:= \left\langle s_1,\cdots, s_N :
\begin{array}{ll}
& s_k^2=e ~{\rm (identity)},~~ (s_{k}s_{k+1})^3=e ~{\rm mod}~N \\
& (s_ks_j)^2=e, ~{\rm if}~k\ne j, j\pm 1~{\rm mod}~N\\
\end{array}
\right\rangle
\end{equation}
Then one can show that the action of $W$ on signs has the following periodic
property:
\begin{Lemma}
\label{cycle}
For each $\sigma\in \Sigma (N)\setminus \{\sigma^0\}$,
there exists a periodic action of length
$N$, $w^{(\sigma)}_N:=s_{i_N}\cdots s_{i_1}\in W$ with $i_j\ne i_k$ if $j\ne
k$,
such that
\begin{itemize}
\item[i)] $w^{\sigma}_N\sigma=\sigma$
\item[ii)] $s_{i_j}\cdots s_{i_1}\sigma\ne s_{i_k}\cdots s_{i_1}\sigma,
\quad {\rm for}
\quad 0\le j <k<N. $
\item[iii)]  N is the minimum number, i.e. if there is $ w(\ne e)\in W$
such that $ w\sigma=\sigma$, then $l(w)\ge N$,
\end{itemize}
where $l(w)$ is the length of the element $w\in W$.
\label{signchange}
\end{Lemma}
\begin{Proof}
Consider a sequence $(\overbrace{+\cdots +}^{n_1}\overbrace{-\cdots -}^{n_2}
\overbrace{+\cdots +}^{n_3}\overbrace{-\cdots -}^{n_4})$ with
$\sum_{k=1}^4n_k=N$,
and assume $\sigma$ be given by its periodic extension of period $N$.
First move the $+$ at the $n_1$-th place to the $(n_1+n_2)$-th place by applying
$w_1=s_{n_1+n_2-1}\cdots s_{n_1+1}s_{n_1}$ with $l(w_1)=n_2$. Also move the
$+$ at the $(n_1+n_2+n_3)$-th place to the $N$-th place by
$w_2=s_{N-1}\cdots s_{n_1+n_2+n_3}$ with $l(w_2)=n_4$. Then we obtain
$(\overbrace{+\cdots +}^{n_1-1}\overbrace{-\cdots -}^{n_2}
\overbrace{+\cdots +}^{n_3}\overbrace{-\cdots -}^{n_4}+)$.
Now move the $-$ at the $n_1$-the place to the $0$-th place by 
$w_3=s_Ns_1\cdots s_{n_1-2}s_{n_1-1}$ with $l(w_3)=n_1$, and
move the $-$ at the $(n_1+n_2+n_3-1)$-th place to the $(n_1+n_2)$-th place by
$w_4=s_{n_1+n_2+1}\cdots s_{n_1+n_2+n_3}$ with $l(w_4)=n_3$.
Thus applying $w=w_4w_3w_2w_1$ with $l(w)=N$, the sequence $\sigma$
returns to the original one. Notice that the sequences in the
middle steps are all different.  It is easy to extend the procedure
to the general case.
\end{Proof}

Let us consider the sign changes in $a_k$'s through a blow-up,
which is determined by $\tau_k(t_*)=0$ for some $k$ and $t_*\in{\Bbb R}$.
Then from (\ref{solutionabk}), if $\tau_k(t) \sim (t-t_*)$ near $t=t_*$,
the solutions $a_k(t)$
and $b_k(t))$ near $t=t_*$ has the following form,
\begin{equation}
\nonumber
\begin{array}{lll}
a_k(t)\sim\displaystyle{{1\over (t-t_*)^2}, }& a_{k-1}(t)\sim
(t-t_*),
& a_{k+1}(t)\sim (t-t_*), \\
b_{k}(t)\sim\displaystyle{{1\over t-t_*},} &
\displaystyle{b_{k+1}(t)\sim{1\over t-t_*}, }& \\
\end{array}
\end{equation}
 and all the other variables
$a_j$ and $b_j$ are regular near $t=t_*$. Observe that the $a_k$ remains the
same sign (negative)
before and after the blow-up point, and $a_{k\pm 1}$ change their signs.
This implies that the sign changes of $a_{k\pm 1}$ through the blow-up of $a_k$
correspond to the permutation $s_k$. Recall $\epsilon_k={\rm sign}(a_k)
=\sigma_k\sigma_{k+1}$. Then Lemma \ref{signchange} suggests
that the chain of $\sigma$'s appearing in the periodic permutation
$w_N^{(\sigma)}$ provides the labels of the cylinders in each torus
$({\Bbb T}^g)_j$,
and the gluing rule is given by the sign-representation of the affine
Weyl group $W$. It may be convenient to use the $\epsilon$-sign for the
label. Then we have the general formula for the sign change in terms of
the extended Cartan matrix $(C_{i,j})_{1\le i,j\le N}$ for $A_{N-1}^{(1)}$,
\begin{equation}
s_k: \epsilon_j \longmapsto \epsilon_j\epsilon_k^{-C_{k,j}}
\end{equation}
which was also obtaind in \cite{casian:99} for the case of the nonperiodic
Toda lattice.
We summarize those as the following Conjecture:
\begin{Conjecture}
\label{conjecture1}
Each connected piece ({\it cylinder}) in the affine parts of the isospectral
level set $Z_{\Bbb R}^P(\gamma)$ can be marked by the signs $(\epsilon_1,
\cdots,\epsilon_N)$ with $\epsilon_k=sign(a_k)$, and they are glued together
according to the action of a length $N$-element of the affine Weyl group $W$
on those signs ({\it sign-representation of W}).
\end{Conjecture}

\begin{Example}
\label{n3pToda} $N=3$ periodic Toda lattice \cite{audin:94}:

\begin{itemize}
\item[i)] $m=2$ ($N_R=3$, three connected components in ${\mathcal R}_{\Bbb
R}$):
The real part of the
Jacobian ${\rm Jac}({\mathcal R})_{\Bbb R}$ consists of {\it four} tori, and
the affine part is given by
\begin{equation}
\nonumber
{\rm Jac}({\mathcal R})_{\Bbb R}\setminus {\Theta}\cong
{\mathcal M}_0^{\circ}\sqcup{\mathcal M}_1^{\circ}\sqcup{\mathcal M}_2^{\circ},
\end{equation}
where
\begin{equation}
\nonumber
\left\{
\begin{array}{ll}
{\mathcal M}_2^{\circ} &=({\Bbb T}^2)_2={\Bbb T}^2 ,\\
{\mathcal M}_1^{\circ} &=({\Bbb T}^2\setminus\Theta)_1\sqcup(
{\Bbb T}^2\setminus\Theta)_1 ,\\
{\mathcal M}_0^{\circ} &=({\Bbb T}^2\setminus\Theta)_0={\Bbb R}^2\sqcup{\Bbb
R}^2\sqcup{\Bbb R}^2 ,
\end{array}
\right. 
\end{equation}
Each $(T^2\setminus\Theta)_1$ consists of {\it three} cylinders ${\Bbb
R}\times S^1$.
The four ${\Bbb T}^2$'s as compactified connected level sets are marked by
the signs
$(\epsilon_1,\epsilon_2,\epsilon_3)$ with $\prod_{k=1}^3\epsilon_k=1$.
For examples, ${\Bbb T}^2$ in the ${\mathcal M}_2^{\circ}$ is marked by
$(+++)$, and
each connected component in $({\Bbb T}^2\setminus\Theta)_j$ for other ${\mathcal M}_j$'s is marked by
either $(+--)$ or $(-+-)$ or $(--+)$. Those components are glued together by
the $W$-action on the signs,
\[
(+,-,-)\overset{s_2}{\longrightarrow}(-,-,+)\overset{s_1}{\longrightarrow}
(-,+,-)\overset{s_3}{\longrightarrow}(+,-,-).
\]

\item[ii)] $m=1$ ($N_R=2$): The ${\rm Jac}({\mathcal
R})_{\Bbb R}$
consists of {\it two} tori, and the affine part is
\begin{equation}
\nonumber
{\rm Jac}({\mathcal R})_{\Bbb R}\setminus {\Theta}\cong
{\mathcal M}_0^{\circ}\sqcup{\mathcal M}_1^{\circ},
\end{equation}
where
\begin{equation}
\nonumber
\left\{
\begin{array}{ll}
{\mathcal M}_1^{\circ}&=({\Bbb T}^2\setminus\Theta)_1=({\Bbb R}\times
S^1)\sqcup
({\Bbb R}\times S^1)\sqcup ({\Bbb R}\times S^1),\\
{\mathcal M}_0^{\circ} &=({\Bbb T}^2\setminus\Theta)_0={\Bbb R}^2\sqcup{\Bbb
R}^2\sqcup{\Bbb R}^2,
\end{array}
\right.
\end{equation}

\item[iii)] $ m=0$ ($N_R=1$): The ${\rm Jac}({\mathcal
R})_{\Bbb R}$
is just one torus, and the affine part is
\begin{equation}
\nonumber
{\rm Jac}({\mathcal R})_{\Bbb R}\setminus {\Theta}\cong
{\mathcal M}_0^{\circ} =({\Bbb T}^2\setminus\Theta)_0={\Bbb R}^2\sqcup{\Bbb
R}^2\sqcup{\Bbb R}^2
\end{equation}
\end{itemize}

\end{Example}

\medskip
\noindent
{\bf Aknowledgement}
The author is grateful to N. Ercolani, H. Flaschka and J. Gibbons
for valuable discussions and suggestions.
He would also like to thank L. Casian who has provided several
useful ideas on the beginning statge of this study 
through their fruitful collaboration.
This work is partially supported by the NSF grant 
 DMS0071523 and
the Newton Institute Program ``Integrable Systems"
for the period of July-September, 2001.

\bibliographystyle{amsplain}

\end{document}